\documentclass[%
preprint,
showkeys,
superscriptaddress,
 prb,
]{revtex4-2}
\usepackage{siunitx}
\usepackage{amsmath}
\usepackage{latexsym}
\usepackage{color}
\usepackage{ulem}
\usepackage{array}
\usepackage{amsmath}
\usepackage[pdftex]{graphicx}
\usepackage{bm}
\usepackage{ascmac}
\usepackage{here}
\usepackage{afterpage}

\begin{document}

"This is the pre-peer reviewed version of the following article: Onoda, H., Sukegawa, H., Inoue, J.-I., Yanagihara, H., Strain Engineering of Magnetic Anisotropy in Epitaxial Films of Cobalt Ferrite. Adv. Mater. Interfaces 2021, 2101034., which has been published in final form at \url{https://doi.org/10.1002/admi.202101034}. This article may be used for non-commercial purposes in accordance with Wiley Terms and Conditions for Use of Self-Archived Versions."

\title{Strain engineering of magnetic anisotropy in epitaxial films of cobalt ferrite}

\author{Hiroshige Onoda} 
\affiliation{Department of Applied Physics, University of Tsukuba, Tsukuba, Ibaraki 305-8577, Japan}

\author{Hiroaki Sukegawa} 
\affiliation{National Institute for Materials Science (NIMS), Tsukuba, Ibaraki 305-0047, Japan}

\author{Jun-ichiro Inoue} 
\affiliation{Department of Applied Physics, University of Tsukuba, Tsukuba, Ibaraki 305-8577, Japan}

\author{Hideto Yanagihara*}
\affiliation{Department of Applied Physics, University of Tsukuba, Tsukuba, Ibaraki 305-8577, Japan}
\affiliation{Tsukuba Research Center for Energy Materials Science (TREMS), University of Tsukuba, Tsukuba, Ibaraki 305-8573, Japan}
\email{yanagihara.hideto.fm@u.tsukuba.ac.jp}

\begin{abstract}
Perpendicular magnetic anisotropy (PMA) energy up to $K_{\mathrm{u}}=6.1\pm\SI{0.8}{MJ.m^{-3}} $ is demonstrated in this study by inducing large lattice-distortion exceeding 3\% at room temperature in epitaxially distorted cobalt ferrite Co$ _{x} $Fe$ _{3-x} $O$ _{4} $ (x = 0.72) (001) thin films. Although the thin film materials include no rare-earth elements or noble metals, the observed $ K_{u} $ is larger than that of the neodymium-iron-boron compounds for high-performance permanent magnets. The large PMA is attributed to the significantly enhanced magneto-elastic effects, which are pronounced in distorted films with epitaxial lattice structures upon introducing a distortion control layer of composition Mg$ _{2-x} $Sn$_{1+x}$O$ _{4} $. Surprisingly, the induced $ K_{u} $ can be quantitatively explained in terms of the agreement between the local crystal field of Co$ ^{2+} $ and the phenomenological magneto-elastic model, indicating that the linear response of induced $K_u$ is sufficiently valid even under lattice distortions as large as 3.2\%. Controlling tetragonal lattice deformation using a non-magnetic spinel layer for ferrites could be a promising protocol for developing materials with large magnetic anisotropies.
\end{abstract}

\keywords{Magnetic anisotropy, Thin films, Magneto-elastic effect, Spinel ferrite, Strain}

\maketitle

\section{Introduction}
Saturation magnetization, Curie temperature, and magnetic anisotropy (MA) are the fundamental properties of ferromagnets. Materials used for fabricating modern permanent magnets require large coercivity and high saturation magnetization to generate sufficient magnetic flux. The mechanism of coercivity is very complicated; unfortunately, no universal model for the coercivity of ferromagnets has yet been reported. However, previous studies have revealed that the maximum value of coercivity is limited by MA.\textsuperscript{\cite{Stoner1948,Kronmuller1981}} In addition, magnetic thin films with large perpendicular magnetic anisotropies (PMAs) are crucial for application in functional spintronic devices and high-density magnetic recording technology.\textsuperscript{\cite{Meng2006,Piramanayagam2007,Ikeda2010,Sbiaa2011,Oh2016}} To explore or synthesize magnetic materials with high coercivity, materials with large MAs are required for obtaining high-performance permanent magnets. Because MA is indirectly proportional to the symmetry of the crystal, materials with lower crystal symmetry are promising for application as high-MA compounds. Generally, the physical origin of large MAs is spin-orbit interaction (SOI), calculated as the inner product of orbital ($\boldsymbol{L}$) and spin angular momenta ($ \boldsymbol{S} $) with an interaction constant $ \lambda $. For a given spin moment, the magnitude of SOI is determined by $ \lambda $ and the expected value of the orbital angular momentum. Magnetic materials composed of noble metals often exhibit large MAs, attributed to the large $ \lambda $ values of noble metals. In contrast, the large MAs of permanent magnets composed of rare-earth elements are attributed to their large values of $ \lambda $ and orbital angular momenta.\textsuperscript{\cite{Meng2006,Sagawa1984,Radu2012,Carcia1985,Shima2002,Sayama2004,Sagawa1985}}

In transition metal alloys or compounds that do not contain heavy or rare-earth elements, $\boldsymbol{L}$ is often quenched in a crystal field. Consequently, magnetic materials with large MAs are rare. In contrast, a certain amount of orbital angular momentum is sometimes retained in oxides, owing to the localized character of the wave functions of transition metal ions in the crystal field. The magnitude of the orbital angular momentum is influenced by the electronic configurations of the magnetic ions; therefore, the MAs of magnetic oxides can be induced/enhanced by introducing asymmetry, such as lattice deformations. This phenomenon can be considered to be magneto-elastic in nature because the change in magnetic state is induced by lattice deformation. Furthermore, a large uniaxial MA can be realized by uniaxial lattice deformation. The epitaxial distortion arising from the lattice mismatch between oxide thin films and their substrates can be employed to effectively induce lattice distortion. 

Co$ _{x} $Fe$ _{3-x} $O$ _{4} $ (CFO) has a cubic lattice, as shown in \textbf{Figure \ref{figure_RHEED_TEM}(a)}, and exhibits a large cubic MA with a N\'{e}el temperature of \SI{769}{\kelvin} for $ x $ = 2.0 and has been reportedly used as permanent magnets.\textsuperscript{\cite{Iida1956}} Extensive magneto-elastic effects have been reported, along with the existence of a large orbital moment in Co$ ^{2+} $.\textsuperscript{\cite{Dorsey1996,Chambers2002,Huang2006,Lisfi2007,Yanagihara2011,Gatel2013,Eskandari2017,Pham2017,Niizeki2013,Tainosho2019,Onoda2018,Onoda2020}} The large cubic MA of bulk CFO has been elucidated theoretically using a single-ion model; the cubic and local trigonal lattice symmetries split the down-spin $ t_{2g} $ state into a singly occupied and doubly degenerate state with a magnetic quantum number $ \ell \sim\pm1 $; moreover, the degeneracy is lifted by SOI.\textsuperscript{\cite{Bozorth1955,Slonczewski1958}}
Once uniaxial lattice distortion is introduced as shown in \textbf{Figure \ref{figure_RHEED_TEM}(b)}, a large PMA should arise because of the magneto-elastic effects. This can be inferred from a phenomenological magneto-elastic model that predicts that induced MA is proportional to the lattice distortion. 

Recently, Tainosho \textit{et al.} observed that CFO films grown on an MgAl$ _{2} $O$ _{4} $ (001) substrate suffering 3.6\% in-plane compressive stress exhibit a large negative $ K_u $ of \SI{-5}{MJ.m^{-3}}, which was confirmed by magneto-torque measurements. The induced $K_{\mathrm{u}}$ could be quantitatively explained by the phenomenological magneto-elastic effects, despite the relatively large distortion of 3.6\%.\textsuperscript{\cite{Tainosho2019}} They also reported the occurrence of lattice relaxation caused by misfit dislocations in the early stages of growth because of the large lattice mismatch between the films and substrates. 
Based on these results, further enhancement of $K_{\mathrm{u}}$ can be expected by inducing a 3--4\% tensile distortion. Moreover, the introduction of uniformity is desirable in cases where the lattice mismatch with the substrate is large. We have recently found that Mg$ _{2} $SnO$ _{4} $(001) (MSO), an oxide with a spinel structure, is suitable for the application of large tensile stresses in CFO thin films.\textsuperscript{\cite{Onoda2018, Onoda2020}} In this study, we attempted to quantitatively control the lattice distortion of CFO thin films to induce large positive $ K_{\mathrm{u}} $ values in the absence of platinum group or rare-earth elements. Two methods were adopted to introduce the distortion into CFO thin films on MSO buffer layers: control of the CFO thickness and control of the MSO lattice constants. Notably, both the methods produce significantly large positive $ K_{\mathrm{u}} $ values (perpendicular MA). The latter method is preferable over the former one for arbitrary control of the uniform distortion of the thin films. We confirmed the largest value of the $ K_{\mathrm{u}} $ to be \SI{6.1}{MJ.m^{-3}}, consistent with that expected from the phenomenological magneto-elastic effects.

\section{Results and Discussion}
\begin{figure}[t]
	\begin{center}
		\includegraphics[width=\linewidth]{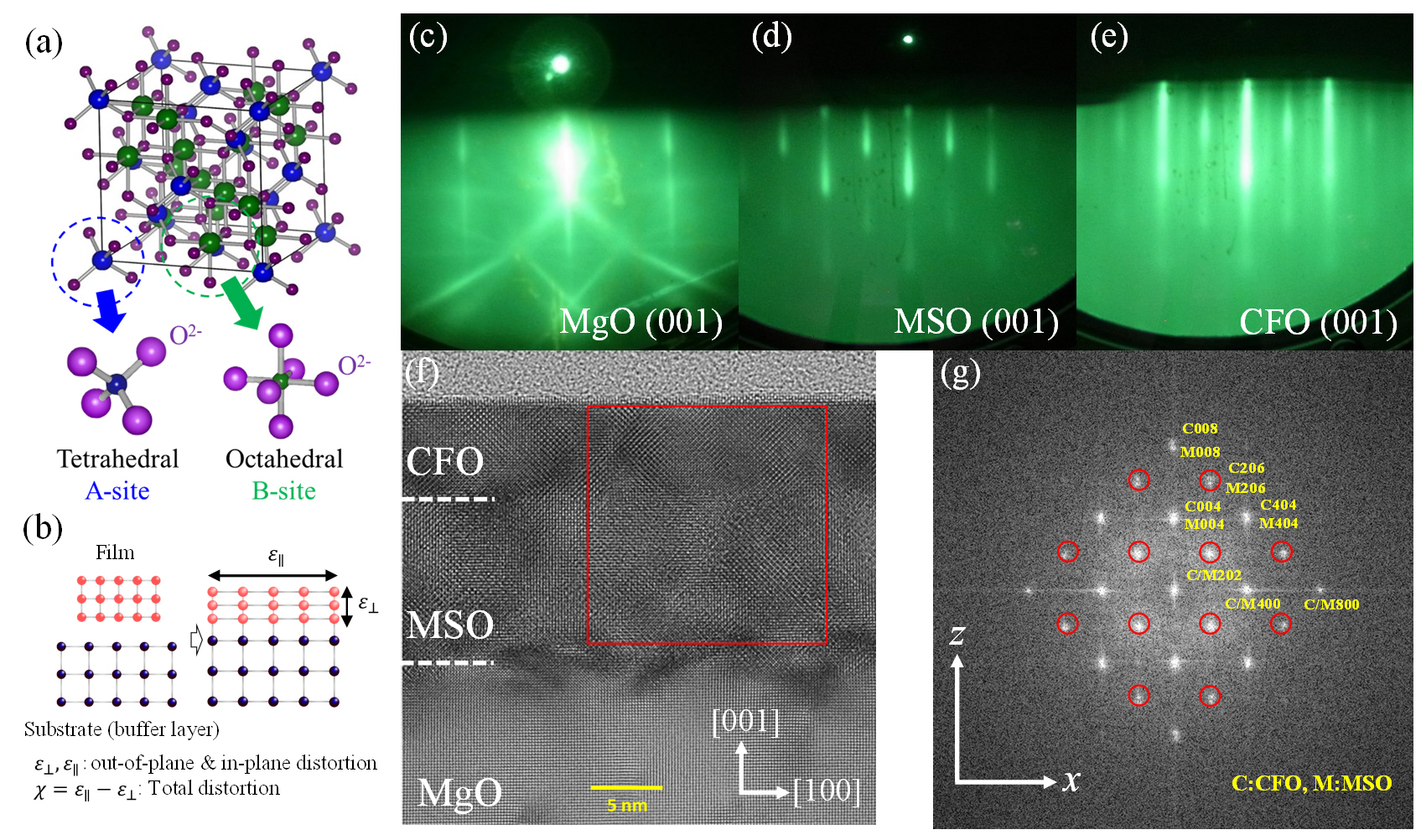}
		\caption{RHEED images and cross-sectional TEM images of CFO/MSO. \textbf{(a)} Spinel structure. \textbf{(b)} Epitaxial distortion and the determination of total distortion $\chi$. RHEED images in [100] azimuth of the \textbf{(c)} MgO substrate, \textbf{(d)} Mg$_2$SnO$_4$ (MSO), and \textbf{(e)} Co-ferrite thin films (CFO). \textbf{(f)} Cross-sectional TEM image. \textbf{(g)} FFT image corresponding to the red square region in \textbf{(d)}. The FFT pattern shows fundamental spots and superlattice spots (red circles) due to the diamond glide of the spinel space group ($ Fd\overline{3}m $).}
		\label{figure_RHEED_TEM}		
	\end{center}
\end{figure}

\textbf{Figure \ref{figure_RHEED_TEM}}(c-e) show the representative reflection high-energy electron diffraction (RHEED) patterns of MgO substrates, post-annealed MSO thin films, and CFO thin films in [100] azimuth. The RHEED patterns of MSO and CFO thin films exhibit typical diffraction patterns from the (001) plane of the spinel structure. The RHEED patterns of the post-annealed MSO thin films displayed a spot-like pattern, whereas that of the CFO displayed a streak-like appearance. \textbf{Figure \ref{figure_RHEED_TEM}}(f) shows the cross-sectional transmission electron microscopy (TEM) images. The interface between the MSO thin films and MgO substrate is rough; this is because post-annealing was performed after the MSO thin film deposition. In contrast, the interface between the CFO and MSO thin films appears flatter compared to the interface of the MSO thin films and MgO substrate. \textbf{Figure \ref{figure_RHEED_TEM}}(g) shows a fast Fourier transform (FFT) image corresponding to the red region in \textbf{Figure \ref{figure_RHEED_TEM}}(f); $ x- $ and $ z- $ denote the in-plane and the out-of-plane directions, respectively. An FFT pattern constitutes fundamental and superlattice spots (red circles in \textbf{Figure \ref{figure_RHEED_TEM}}(f)) owing to the diamond glide of the spinel space group ($ Fd\overline{3}m $).\textsuperscript{\cite{SickafusS2003}}
This result is consistent with the RHEED results. As shown in \textbf{Figure \ref{figure_RHEED_TEM}}(g), the spots are split into two along the $ z-$ direction, implying that the CFO and MSO thin films have identicalin-plane lattices and mismatched out-of-plane lattices. This indicates that epitaxial distortion is introduced in the CFO thin film.

\begin{figure}[t]
	\begin{center}
		\includegraphics[width=\linewidth]{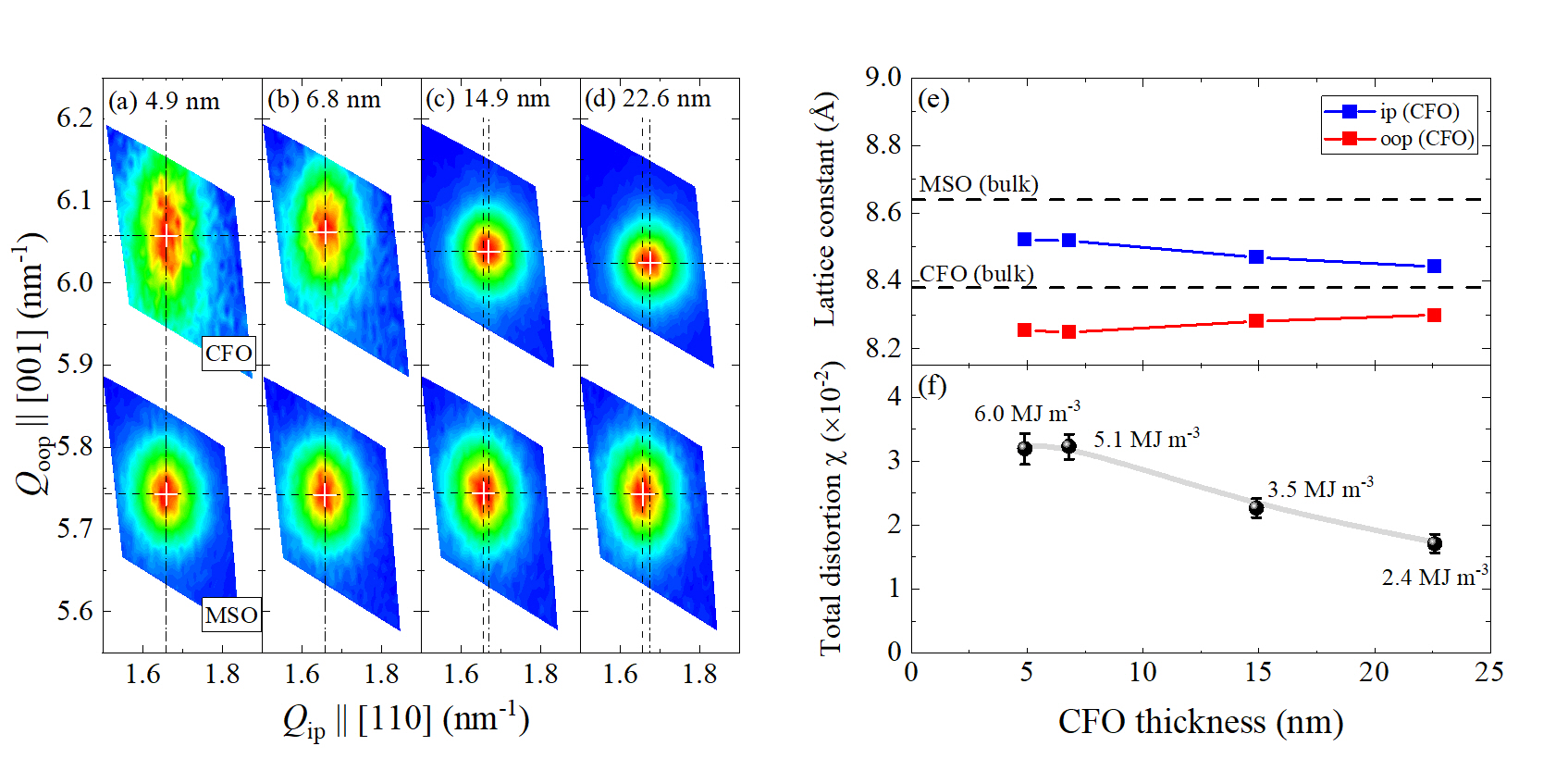}
		\caption{Structural properties of CFO/MSO prepared through Method A. Reciprocal space map of Co-ferrite (CFO) and (Mg, Sn)$ _{3} $O$ _{4} $ (MSO) (115). The thicknesses of the CFO thin films are \textbf{(a)} \SI{4.9}{nm}, \textbf{(b)} \SI{6.8}{nm}, \textbf{(d)} \SI{14.9}{nm}, and \textbf{(d)} \SI{22.6}{nm}. The black dashed and chained lines represent Q$ _{ip} $, which are the same for CFO and MSO thin films. This implies that the  CFO thin films are epitaxially locked on to the MSO thin films. \textbf{(e, f)} CFO-thickness-dependence of in-plane and out-of-plane lattice constants and the distortion of the CFO thin films.}
		\label{figure_RSM_MethodA}		
	\end{center}
\end{figure}

\begin{figure}[t]
	\begin{center}
		\includegraphics[width=\linewidth]{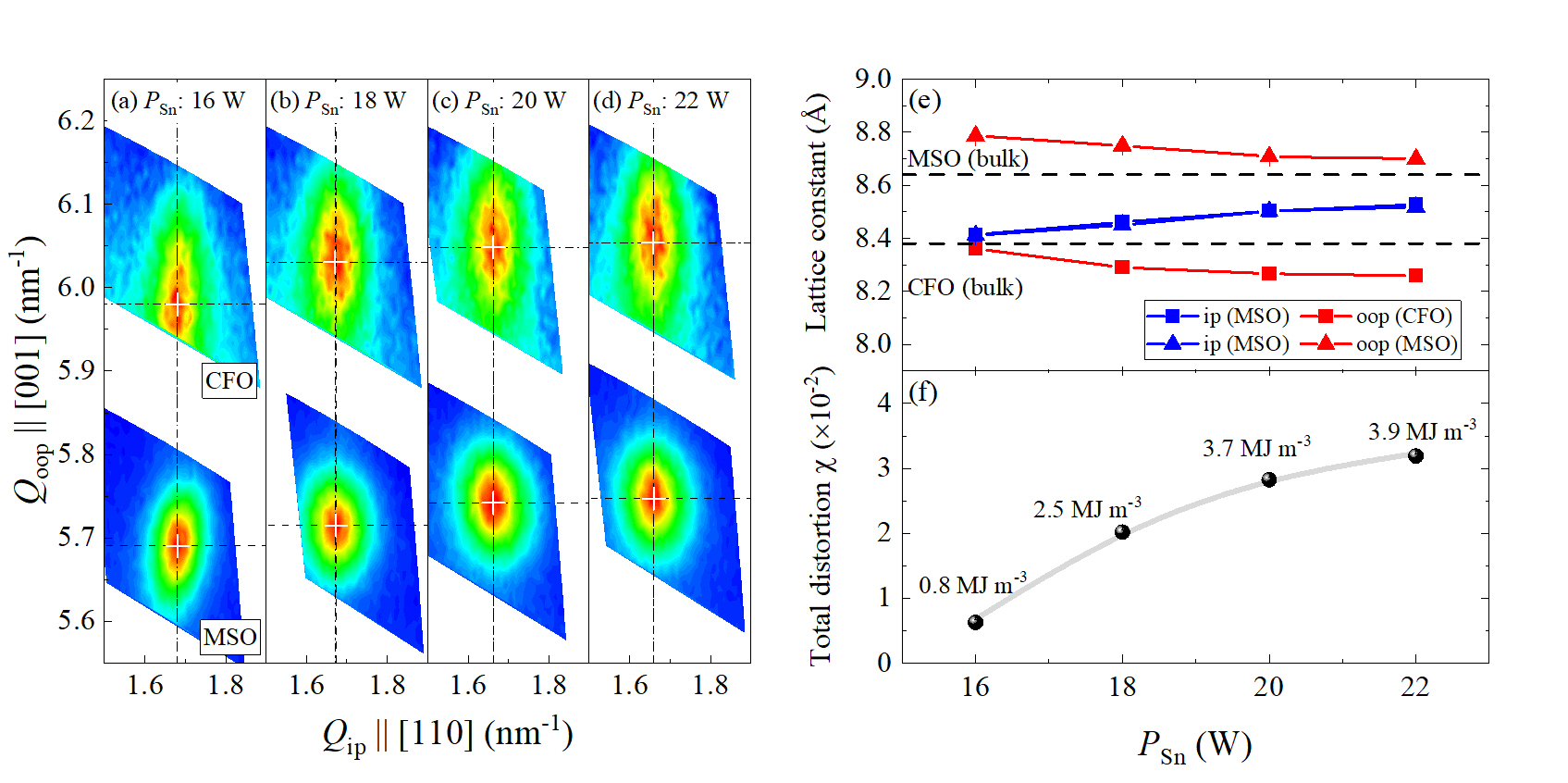}
		\caption{Structural properties of CFO/MSO. \textbf{(a-d)} Reciprocal space map of Co-ferrite (CFO) and (Mg, Sn)$ _{3} $O$ _{4} $ (MSO) (115) prepared through Method B. The thicknesses of the CFO and MSO thin films are \SI{5}{nm} and \SI{10}{nm}, respectively. Q$ _{ip} $ is the same for CFO and MSO thin films, implying that the CFO thin films are epitaxially locked on to the MSO thin films. \textbf{(e)} Input power into the Sn target ($ P_{\mathrm{Sn}} $)-dependence of the in-plane and out-of-plane lattice constants of the CFO and MSO thin films. \textbf{(f)} $ P_{\mathrm{Sn}} $-dependence of the CFO distortion.}
		\label{figure_RSM_MethodB}		
	\end{center}
\end{figure}

To determine the lattice constants of the distorted thin films, reciprocal space map (RSM) measurements were performed. \textbf{Figure \ref{figure_RSM_MethodA}} and \textbf{\ref{figure_RSM_MethodB}}(a-d) display the RSM around the CFO and MSO (115) prepared through Methods A (CFO thickness control)and B (MSO composition control), respectively. The vertical and horizontal axes represent the reciprocal lattice vectors parallel to the out-of-plane and in-plane directions, respectively. The diffraction peaks of the thin films can be obtained by fitting with a 2D Gaussian function. In the samples prepared through Method A, when the CFO thin films are thinner than \SI{10}{nm}, the peak positions of the CFO and MSO along the in-plane direction are equal within the experimental error. This indicates that the CFO thin films are epitaxially locked to the MSO lattice. When the thickness of the thin films exceeds \SI{10}{nm}, lattice relaxation occurs, and the lattice constants advance to the bulk value.
Contrastingly, in the samples prepared through Method B, because the thickness of the CFO thin films is \SI{5}{nm} throughout the entire $P_{\mathrm{Sn}}$, the peak positions of the CFO and MSO along the in-plane direction are equal within the experimental error. This implies that the CFO thin films are epitaxially locked to the MSO lattice. The $ P_{\mathrm{Sn}} $-dependence of the lattice constants of the thin films are shown in \textbf{Figure \ref{figure_RSM_MethodB}}(e). The in-plane lattice constants of the CFO and MSO thin films are identical for each $ P_{\mathrm{Sn}} $. The lattice constant of the MSO thin films advance to the bulk value (\SI{8.64}{\angstrom}) as the $ P_{\mathrm{Sn}} $ increases. This implies that the distortion introduced into the CFO thin films increases as the $ P_{\mathrm{Sn}} $ increases.
The in-plane, out-of-plane, and the total distortion of the CFO thin films are defined as $\varepsilon _{\mathrm{ip}} = (a_{\mathrm{ip}}-a_0)/a_0$, $\varepsilon_{\mathrm{oop}} = (a_{\mathrm{oop}}-a_0)/a_0$, and $\chi=\varepsilon _{\mathrm{ip}}-\varepsilon _{\mathrm{oop}}$, respectively. Here, $ a_{\mathrm{ip}} $, $ a_{\mathrm{oop}} $, and $ a_0 $ are the lattice constants of the in-plane, out-of-plane, and bulk CFO ($ a_0 $ = \SI{8.38}{\angstrom}), respectively. 
The total distortion $ \chi $ decreases as CFO thickness increases, as is evident from \textbf{Figure \ref{figure_RSM_MethodA}}(f). Similarly, $ \chi $ also increases as $ P_{\mathrm{Sn}} $ increases, as seen in\textbf{ Figure \ref{figure_RSM_MethodB}}(f). 
The maximum $\chi$ is 0.032 in both the methods and is twice as large compared to the value when the deposition occurs on the MgO substrates.\textsuperscript{\cite{Niizeki2013}} According to the magneto-elastic effect, the induced uniaxial MA $ K_\mathrm{u} $ can be expressed as $ K_\mathrm{u} = B_1 \chi $. Here, $ B_1 $ is the magneto-elastic constant, which is \SI{0.14}{GJ.m^{-3}} for bulk CFO ($ B_1=3/2 \lambda_{100} (C_{12} - C_{11})$, where $ \lambda_{100} = -590\times10^{-6} $, $ C_{11} $ = \SI{273}{GPa}, $ C_{12} $ = \SI{106}{GPa}).\textsuperscript{\cite{Bozorth1955,Schulz1994,Hu2000}} Therefore, we can estimate that the induced $ K_\mathrm{u} $ is approximately \SI{4.5}{MJ.m^{-3}} for 3.2\% distorted CFO thin films.

\begin{figure}[t]
	\begin{center}
		\includegraphics[width=\linewidth]{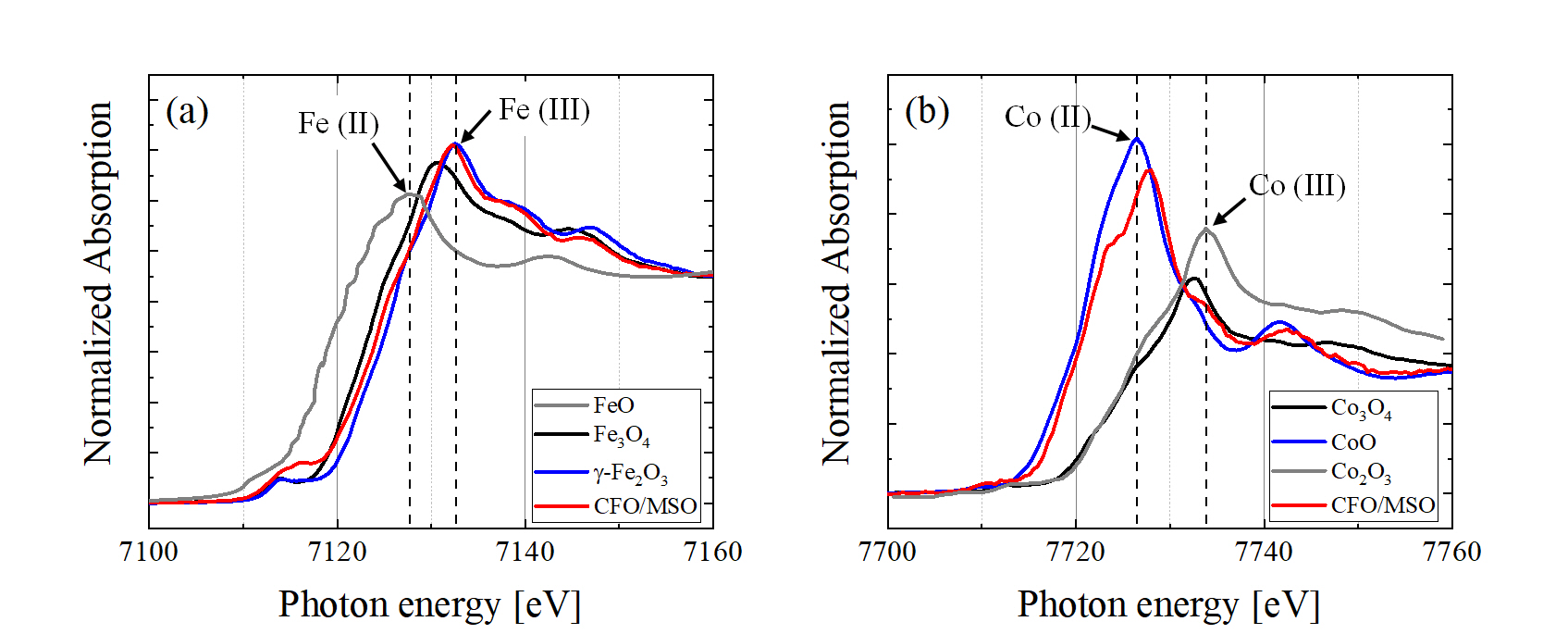}
		\caption{K-edge XANES spectra of CFO/MSO. \textbf{(a)} Fe K-edge XANES of $ \gamma $-Fe$ _{2} $O$ _{3} $, Fe$_3$O$_4$, and MSO/CFO thin films. The CFO on the MSO film is similar to that of $ \gamma $-Fe$ _{2} $O$ _{3} $. \textbf{(b)} Co K-edge XANES of Co$_3$O$_4$, CoO and MSO/CFO thin films. The spectrum of CFO on MSO is similar to that of CoO. ($ \gamma $-Fe$ _{2} $O$ _{3} $, Fe$_3$O$_4$, Co$_3$O$_4$, and CoO were prepared in our laboratory.)}
		\label{figure4}		
	\end{center}
\end{figure}

The valence state of the cations in the CFO thin films were investigated using X-ray absorption near edge structure (XANES) spectra for Fe and Co K-edges, as demonstrated in \textbf{Figure \ref{figure4}}(a) and (b), respectively. For comparison, the spectra of FeO\textsuperscript{\cite{Rubio-Zuazo2018}} in the presence of Fe$ ^{2+} $, Fe$ _{3} $O$ _{4} $ in the presence of Fe$ ^{2+} $ and Fe$ ^{3+} $, and $ \gamma $-Fe$ _{2} $O$ _{3} $ in the presence of Fe$ ^{3+} $, are included in \textbf{Figure \ref{figure4}}(a). Additionally, the spectra of CoO in the presence of Co$ ^{2+} $, Co$ _{3} $O$ _{4} $ in the presence of Co$ ^{2+} $ and Co$ ^{3+} $, and Co$ _{2} $O$ _{3} $\textsuperscript{\cite{Takahashi2002}} in the presence of Co$ ^{3+} $, are shown in \textbf{Figure \ref{figure4}}(b). Furthermore, except for FeO and Co$ _{2} $O$ _{3} $, all samples were prepared in our laboratory. The spectrum of Fe in the CFO film is close to that of $ \gamma $-Fe$ _{2} $O$ _{3} $, indicating that Fe$ ^{3+} $ is dominant, and the vacancies at the octahedral sites exist as in $ \gamma $-Fe$ _{2} $O$ _{3} $ \cite{Yanagihara2006} owing to the Fe-rich composition of the CFO film ($ x = 0.72 $). In the XANES spectrum of Co in the CFO films, the strongest peak of the CFO thin films appeared at approximately \SI{7728}{eV}, corresponding to Co$ ^{2+} $; however, the peak position was slightly shifted to the high-energy region due to the difference in structure. The structure of the CoO is rock salt, while the CFO exhibits the spinel structure. Therefore, we can conclude that the valence of Co in the CFO thin films was Co$ ^{2+} $. Consequently, the cations are distributed as [Fe$ ^{3+} $]$ _{\mathrm{A}} $(Co$ ^{2+}_{8/11} $Fe$ ^{3+}_{13/11} $$ \Box $$ _{1/11} $)$ _{\mathrm{B}} $O$ _{4} $ in the CFO thin films. Here, [ ]$ _{\mathrm{A}} $ and ( )$ _{\mathrm{B}} $ represent the A-sites and B-sites of the spinel structure, respectively, and $ \Box $ represents the vacancies.

\begin{figure}[t]
	\begin{center}
		\includegraphics[width=\linewidth]{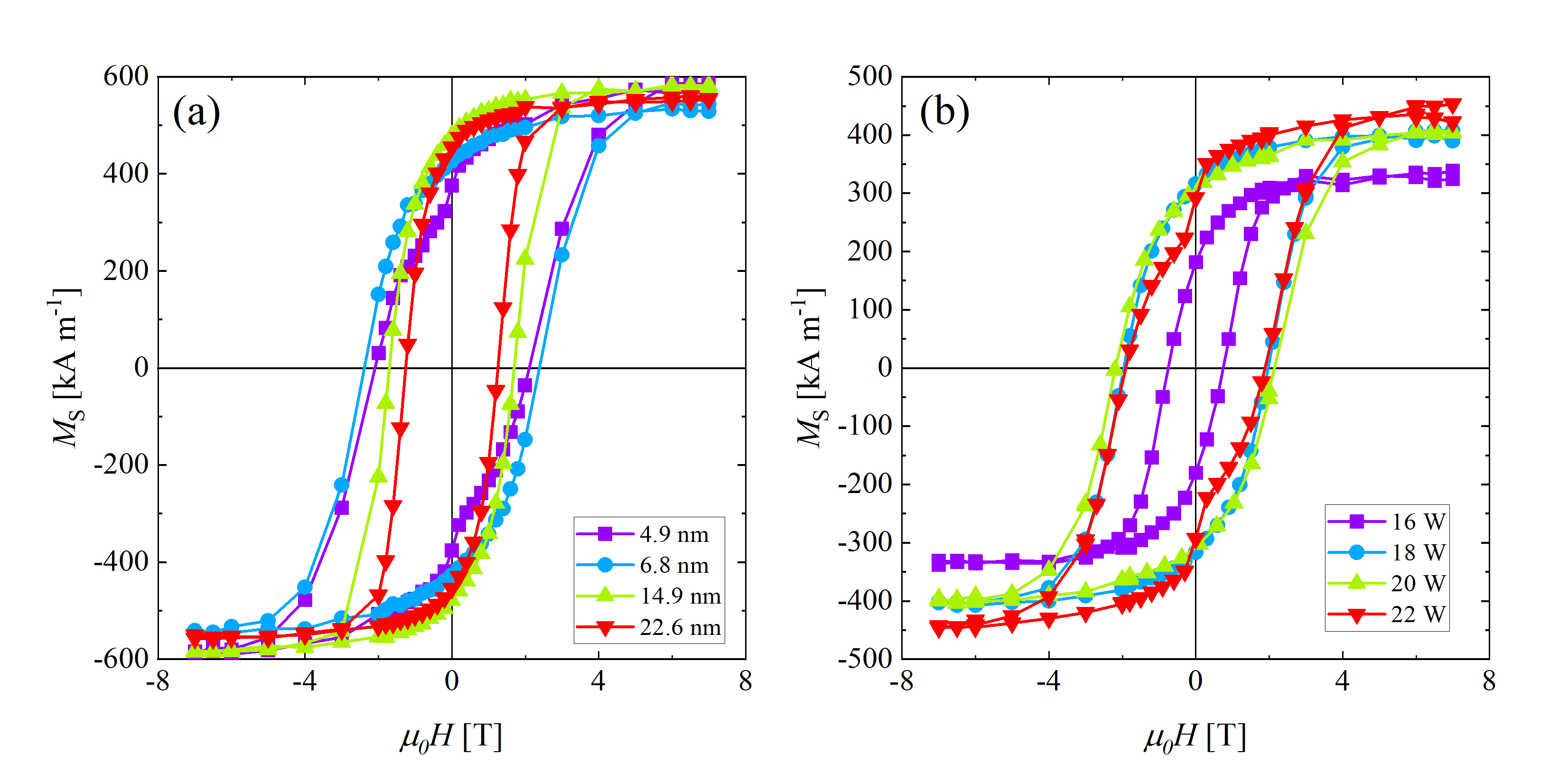}
		\caption{Magnetization processes of CFO/MSO. The out-of-plane magnetization curves of the CFO thin films prepared in \textbf{(a)} Method A and \textbf{(b)} Method B at room temperature. The maximum applied field was $\pm$\SI{7}{\tesla}.}
		\label{figure5}		
	\end{center}
\end{figure}

\textbf{Figure \ref{figure5}} displays the out-of-plane $MH$ curves of the CFO thin films at room temperature. The maximum applied field was $\pm$\SI{7}{\tesla}.
In both the methods, the coercivity tends to increase as the introduced distortion increases. 
All the values of the saturation magnetization $M_\mathrm{S}$ of the CFO thin films were comparable to or larger than the bulk value (\SI{425}{kA.m^{-1}}).\textsuperscript{\cite{Cullity2011}}

\begin{figure}[t]
	\begin{center}
		\includegraphics[width=\linewidth]{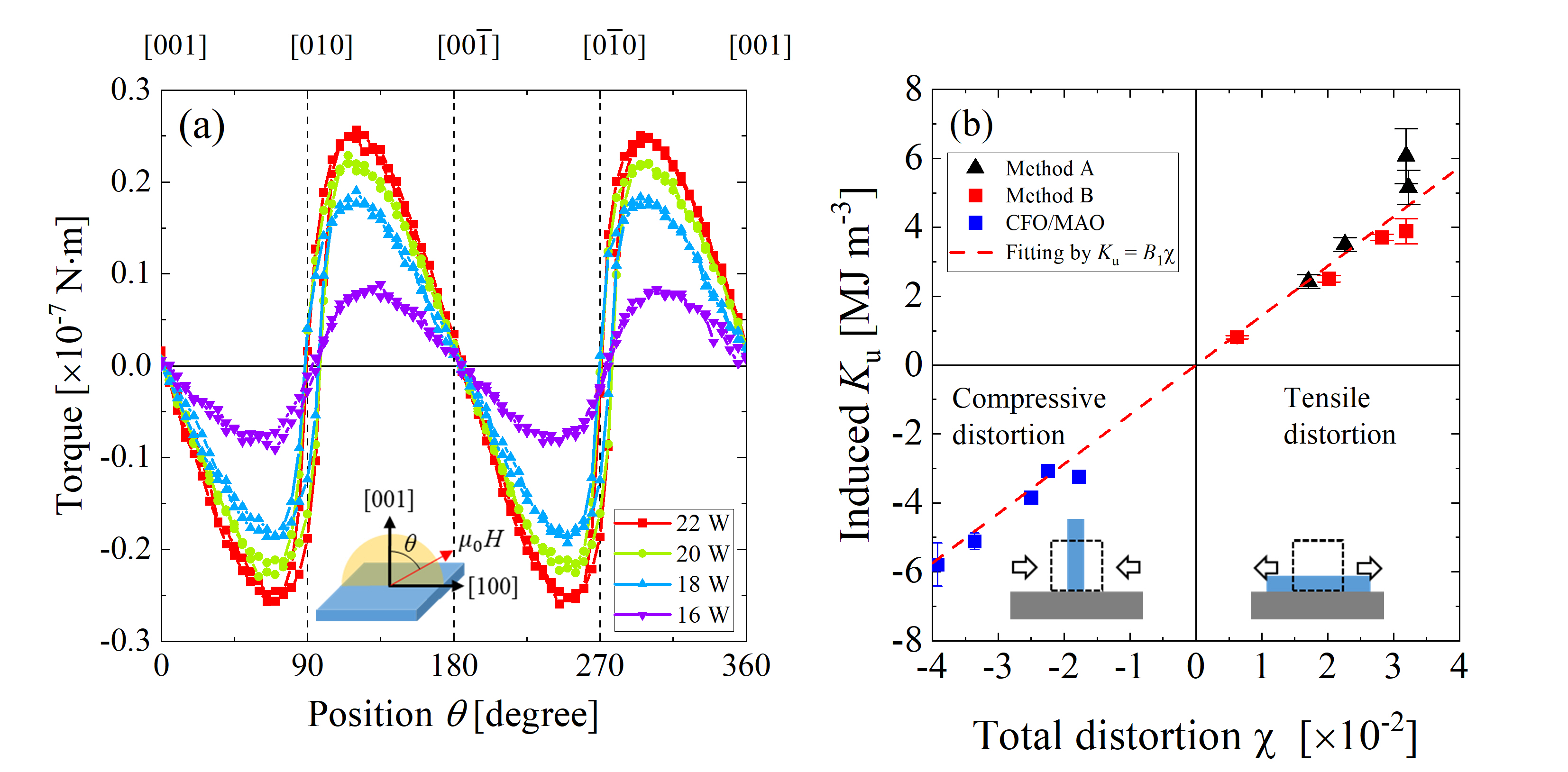}
		\caption{ Magnetic anisotropy of CFO/MSO. \textbf{(a)} Torque measurement curves applied $\mu_{0}H=\SI{9}{\tesla}$ for CFO/MSO thin films at room temperature. \textbf{(b)} Total distortion $ \chi $-dependence of the induced magnetic anisotropy $ K_\mathrm{u} $. The straight line fits the result in the form of a linear function. The magneto-elastic constant can be obtained from the slope as $ B_1 = 0.148 \pm \SI{0.006}{GJ.m^{-3}} $.}
		\label{figure6}		
	\end{center}
\end{figure}

The values of the MAs were determined by magnetic torque measurements. \textbf{Figure \ref{figure6}}(a) shows the torque curves for the CFO/MSO thin films prepared through Method B under $\mu_{0}H=\SI{9}{\tesla}$. The magnetic torque $ T(\theta) $ around the [001]-direction is represented by 
$T(\theta) = -K_\mathrm{u}^{\mathrm{eff}} \sin2\theta$.
Here, $ \theta $ is the angle between the magnetization and the [001]-direction. As evident from \textbf{Figure \ref{figure6}}(a), the magnitude of the torque curve increased as $ P_{\mathrm{Sn}} $ increased, indicating that the MA increases with an increase in the lattice distortion. Even under an applied field of \SI{9}{\tesla}, the shapes of all the torque curves deviated  considerably from the sinusoidal curve, and rotational hysteresis was observed at the crossing of the magnetic hard axis. This suggests the presence of an exceedingly large uniaxial MA with a preferential axis normal to the film plane.
Because the anisotropic magnetic field is larger than the applied magnetic field, the intrinsic $ K_\mathrm{u}^{\mathrm{eff} }$ was derived using the extrapolation method. When the magnetic torque $ L $ measured under the applied field \textit{H} is plotted in the direction of \SI{45}{\degree} to the film surface as a function of $ (L/H)^2 $, it becomes linear.\textsuperscript{\cite{Miyajima1976}} 
From the extrapolation, the $ K_\mathrm{u}^{\mathrm{eff}} $ was estimated to be $5.9 \pm \SI{0.8}{MJ.m^{-3}}$ at its maximum. Considering the shape anisotropy of the film form, the MA induced by the epitaxial distortion was $6.1 \pm \SI{0.8}{MJ.m^{-3}}$. This value was considerably large for a ferromagnetic oxide and larger than the value of rare-earth compounds, such as Nd$ _{2} $Fe$ _{14} $B ($\sim \SI{4.9}{MJ.m^{-3}}$).\textsuperscript{\cite{Sagawa1985}} \textbf{Figure \ref{figure6}}(b) shows the total distortion $ \chi $-dependence of the induced $ K_\mathrm{u} $ of all the samples. Although there are variations in $K_{\mathrm{u}}$ for $\chi \sim 0.03$, the $K_{\mathrm{u}}$ values of the samples prepared through either method is proportional to the distortion $\chi$. The slope corresponding to the magneto-elastic constant $ B_1 $ was $0.148 \pm \SI{0.006}{GJ.m^{-3}}$. This value is approximately equal tobulk value of \SI{0.14}{GJ.m^{-3}}, indicating that the phenomenological theory of the magneto-elastic effect is still valid even under the large $ \chi $ value of 0.032. Moreover, it is evident from the results that larger $ K_\mathrm{u} $ can be expected by introducing further distortion.

\section{Conclusion}

CFO epitaxial thin films were grown by RF magnetron sputtering on MSO thin films and the MA induced by lattice distortion was investigated. Two methods were adopted to grow CFO thin films on MSO buffer layers: control of the CFO thickness and control of the MSO lattice constants. We found both the methods to be extremely productive with large positive $ K_{\mathrm{u}} $ values. The induced $ K_u$ was approximately \SI{6}{MJ.m^{-3}} larger than the MA of Nd$ _{2} $Fe$ _{14} $B. Furthermore, the induced $ K_u $ could be explained quantitatively based on the phenomenological magneto-elastic theory, indicating that larger $ K_\mathrm{u} $ could be realized by introducing further distortion. Since a large MA is essential for improving spintronic devices, the strain engineering technique demonstrated in this study could prove promising for the induction of large $K_u$ in the absence of heavy metals or rare-earth elements.

\section{Experimental Section}
The thin films were deposited by reactive RF magnetron sputtering by a multi-cathode system. MSO thin films (thickness 10 nm) were grown at \SI{600}{\degreeCelsius}, employing Mg and Sn metal targets. The thus-grown MSO thin films were post-annealed at \SI{1000}{\degreeCelsius} for \SI{10}{min} in air. CFO thin films were grown on the post-annealed MSO thin films using a CoFe (1:3 atomic ratio) metal alloy. The introduction of distortion in the CFO thin films was attempted by two methods: either by altering the thicknesses of the CFO thin films (Method A), or by changing the lattice constants of the MSO thin films (Method B). In Method A, the CFO thicknesses were varied in the range from 4.9 to \SI{22.6}{nm}. In Method B, the CFO thickness was set to \SI{5}{nm} to prevent lattice relaxation; instead, the lattice constants of MSO were controlled by changing the power input to a Sn target (the power input to the Mg target was constant at \SI{100}{W}, while the power input to the Sn target $P_{\mathrm{Sn}}$ was varied as $P_{\mathrm{Sn}}$ = \numlist{16;18;20;22} \si{\watt}). The surface structure of the thin films was observed by RHEED immediately after growth. The film thickness, determined by X-ray reflectivity (XRR) using a Rigaku Miniflex two-axis X-ray diffractometer, was (Cu K$_{\alpha}$; $\lambda$=\SI{1.5418}{\angstrom}). In-plane and out-of-plane lattice constants of the CFO thin films, determined using a Rigaku SmartLab four-axis X-ray diffractometer, were (Cu K$_{\alpha1}$; $\lambda$=\SI{1.5405}{\angstrom}). All the magnetic measurements were conducted at room temperature. Magnetization was measured using a superconducting quantum interference device, SQUID-VSM (Quantum Design, MPMS) in fields up to $ \pm $\SI{7}{\tesla}. The MAs were evaluated from the magnetic torque curves recorded over the range from zero to \SI{9}{\tesla} using a torque magnetometer (Quantum Design, PPMS Tq-Mag).

\medskip
\textbf{Supporting Information} \par 
Supporting Information is available from the Wiley Online Library or from the author.

\medskip
\textbf{Acknowledgments} \par 
This research was supported by the Japan Science and technology Agency (JST) under Collaborative Research Based on Industrial Demand High Performance Magnets: Towards Innovative Development of Next Generation Magnets (JPMJSK1415). This study was performed with the approval of the Photon Factory Program Advisory Committee (Proposal No. 2015G655, 2017G602, 2016S2-005 and 2019S2-003). H. O. was supported by Grant-in-Aid for JSPS Fellows (19J12384).

\medskip

%


\newpage


\afterpage{\newpage}
\newpage
\clearpage

\textbf{\underline{Supplementary Material}}
\section*{Samples prepared in Method A}
Method A was employed to grown samples twice at different times (Sample A-1 and Sample A-2). The structural parameters and magnetic properties of the fabricated samples are summarized in Table \ref{tb:CFO}.

\begin{table*}[h]
\centering
\caption{Structural parameters and magnetic properties of CFO samples prepared via Method A}
\label{tb:CFO}
\begin{ruledtabular}
\begin{tabular}{lcccccccc}
&Thickness & a & c & $\varepsilon_{\mathrm{ip}}$ & $\varepsilon_{\mathrm{perp}} $ & Total distortion $\chi$ & $M_{\rm S}$ & $K_{\rm u}$ \\ 
&(nm)&(\AA)&(\AA)&($10^{-2}$)&($10^{-2}$)&($(\varepsilon_{\mathrm{ip}}-\varepsilon_{\mathrm{oop}})10^{-2}$)&(kA~m$^{-1}$)&(MJ~m$^{-3}$)\\\hline
Sample A-1&4.9&8.521&8.253&1.68&-1.51&3.19&580&6.06\\
      &6.8&8.518&8.248&1.65&-1.58&3.23&540&5.16\\
      &14.9&8.470&8.280&1.07&-1.19&2.26&580&3.50\\
      &22.6&8.442&8.299&0.740&-0.968&1.71&555&2.43\\ \hline
Sample A-2&5.39&8.492&8.301&1.30&-0.96&2.27&394&2.40\\
       &10.7&8.460&8.278&0.978&-1.21&2.18&441&2.45\\
       &15.8&8.445&8.293&0.8&-1.04&-1.12&497&1.68\\
       &21.0&8.397&8.314&0.23&-0.79&1.02&415&1.44\\
       &40.9&8.366&8.352&-0.18&-0.16&-0.34&439&0.788\\
       &51.2&8.354&8.365&-0.32&-0.19&-0.13&415&0.577\\
\end{tabular}
\end{ruledtabular}
\end{table*}

\newpage

\begin{figure}[h]
    \centering
    \includegraphics[width=0.7\linewidth]{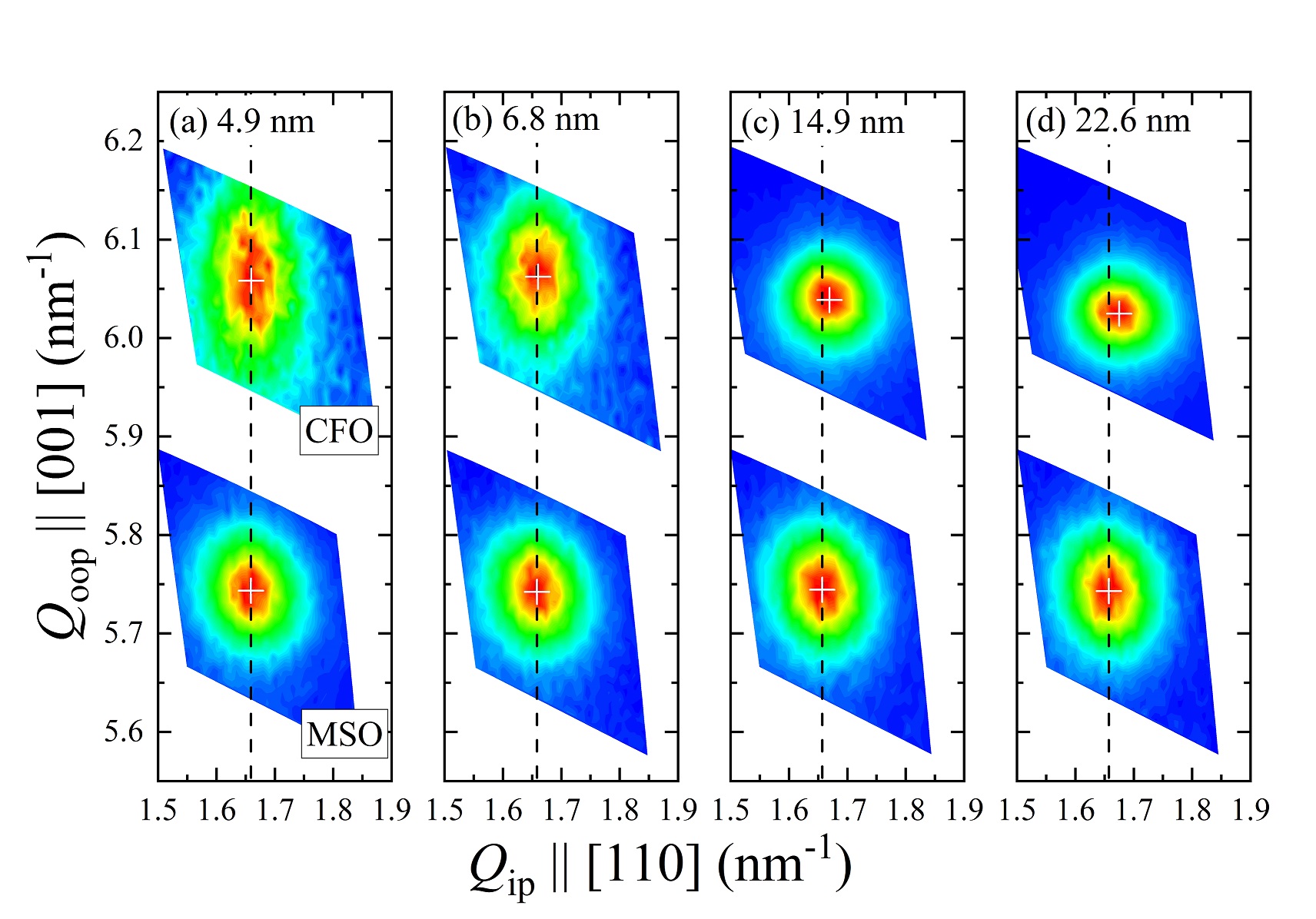}
    \caption{Reciprocal space map of CFO and MSO(115) for CFO samples with thicknesses (a) 5 nm, (b) 7 nm, (c) 15 nm, and (e) 23 nm in Sample A-1.}
    \label{fig:RSM1}
\end{figure}

\begin{figure}[h]
    \centering
    \includegraphics[width=0.7\linewidth]{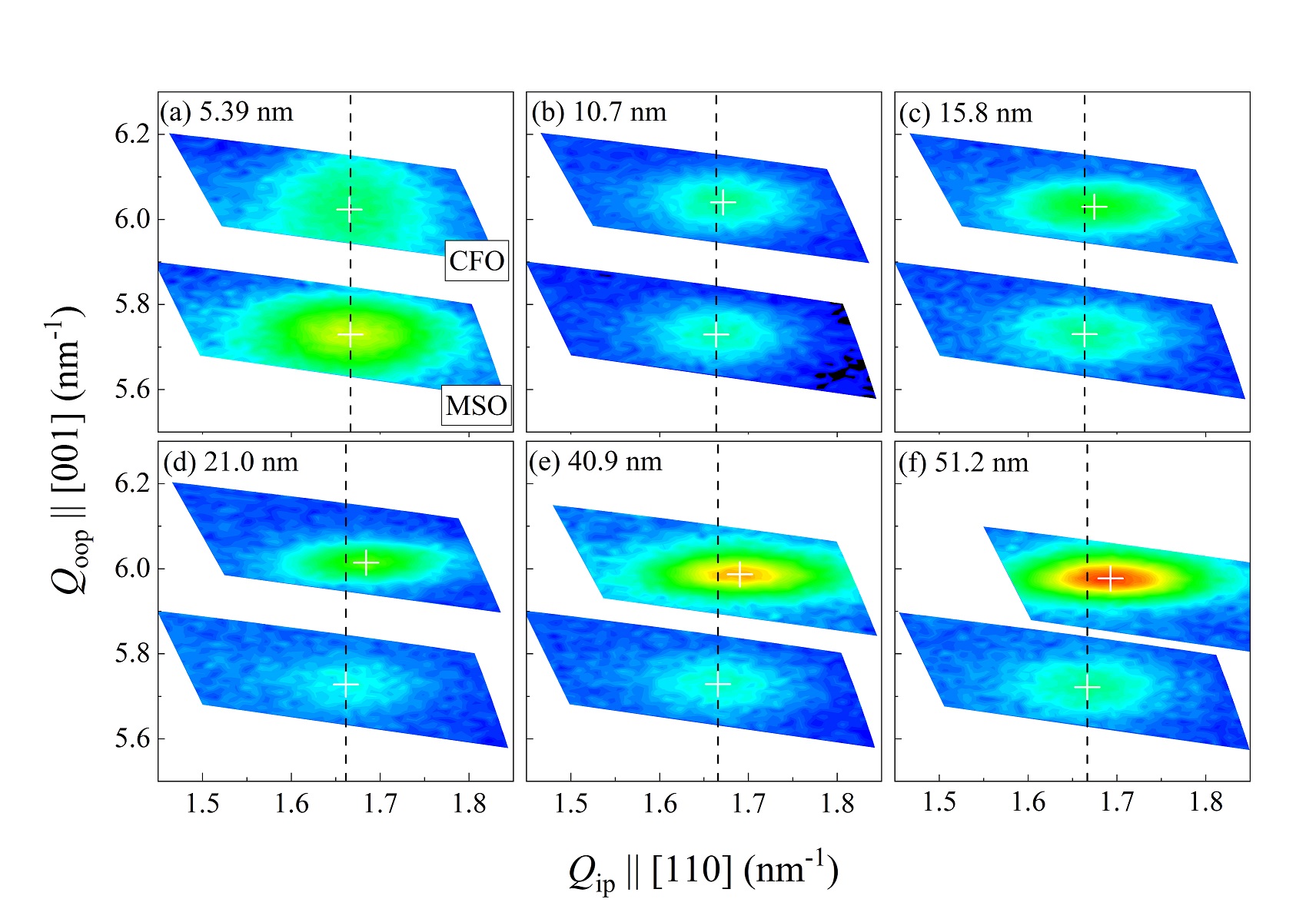}
    \caption{Reciprocal space map of CFO and MSO(115) for CFO samples with thicknesses (a) 5.39 nm, (b) 10.66 nm, (c) 15.76 nm, (d) 40.9 nm, and (e) 51.2 nm in Sample A-2.}
    \label{fig:RSM2}
\end{figure}

\begin{figure}[h]
    \centering
    \includegraphics[width=0.7\linewidth]{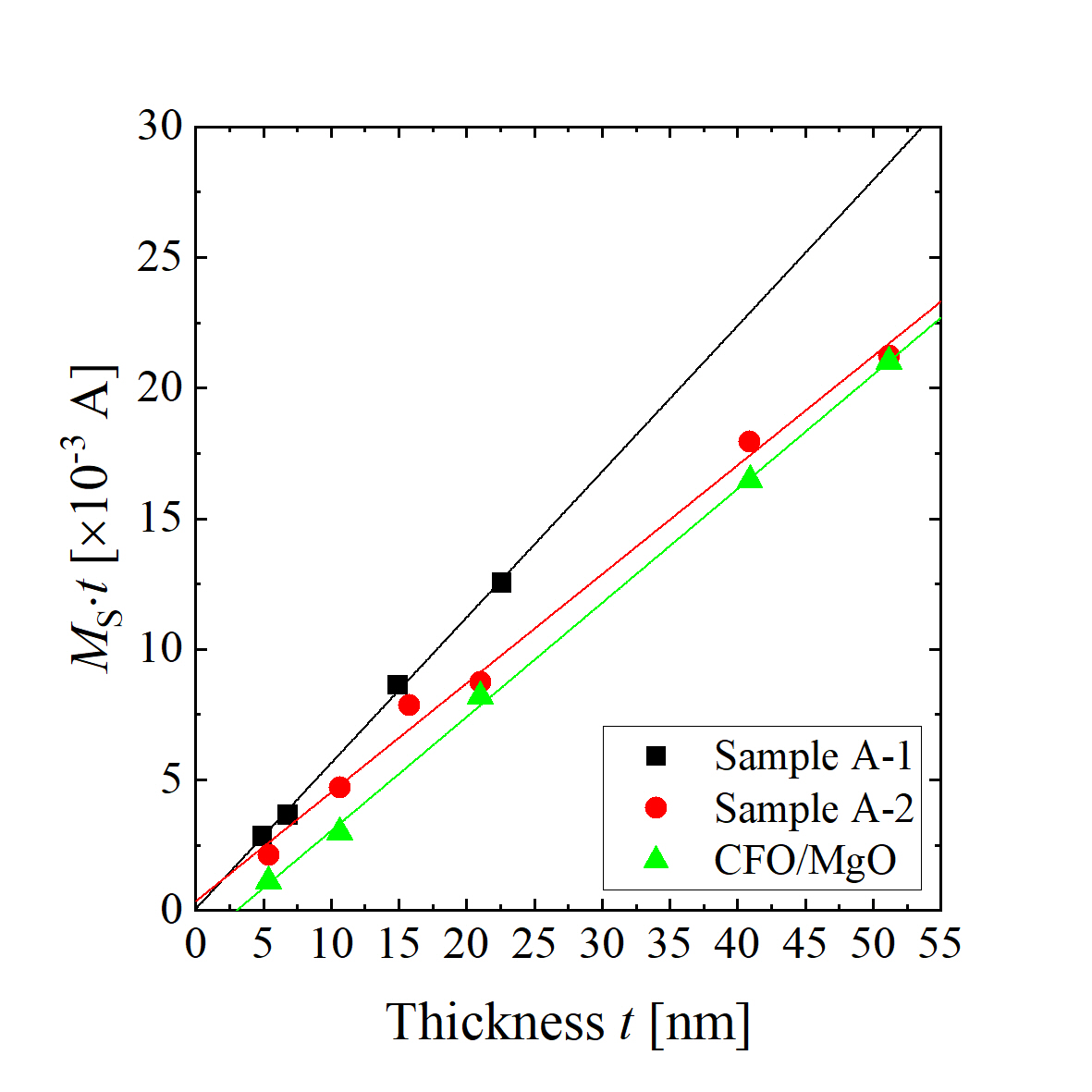}
    \caption{Thickness-dependences of $M_{\mathrm{S}}\cdot t$. }
    \label{fig:Mtplot}
\end{figure}

\begin{figure}[h]
    \centering
    \includegraphics[width=0.7\linewidth]{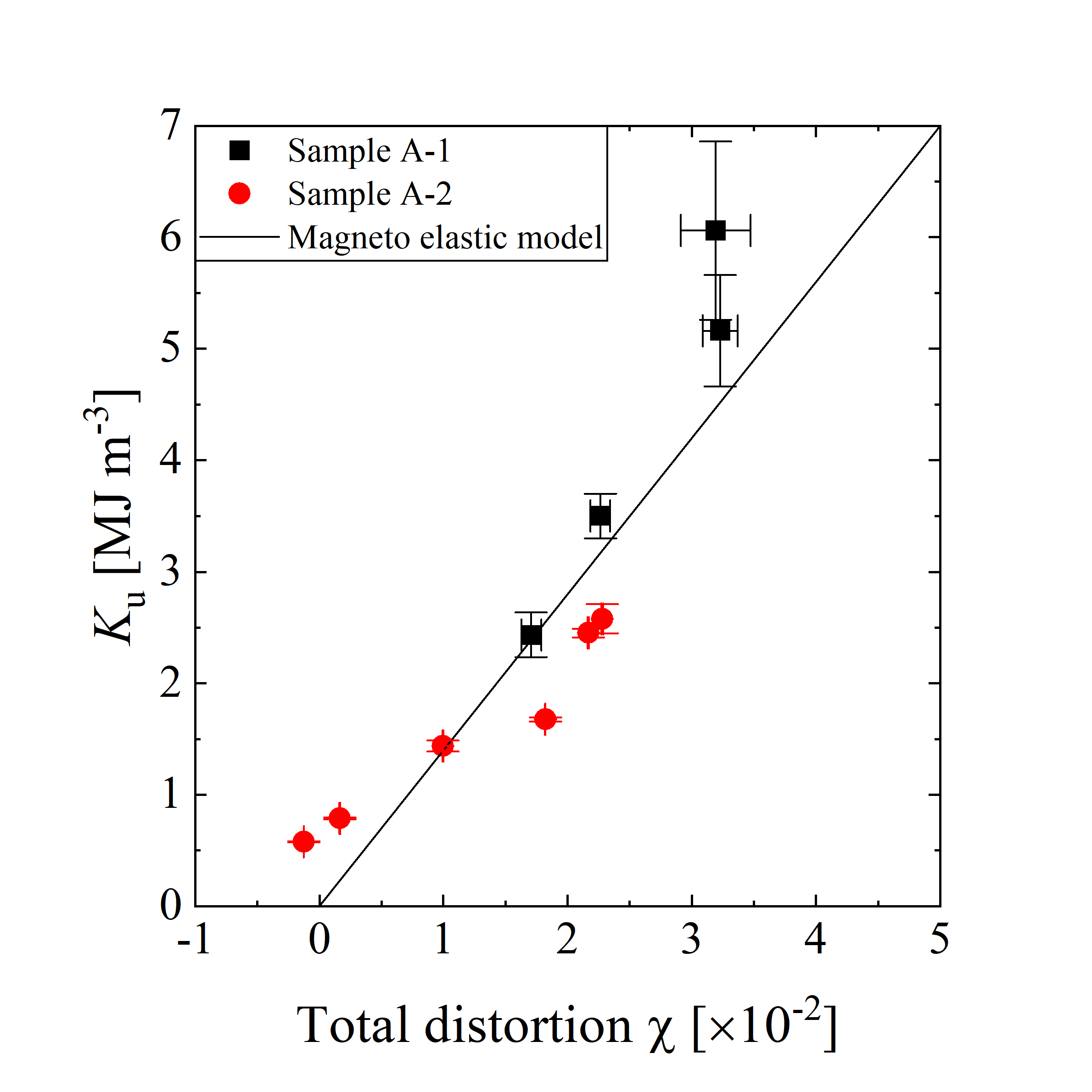}
    \caption{Total distortion $\chi$-dependence of the induced $K_{\mathrm{u}}$.}
    \label{fig:Ku_chi}
\end{figure}

The reciprocal space maps for CFO and MSO(115) are displayed in Figure \ref{fig:RSM1} and Figure \ref{fig:RSM2}, respectively. The vertical and horizontal axes represent the reciprocal lattice vector parallel to the out-of-plane and in-plane directions, respectively. The diffraction peak positions for each thin film were obtained by fitting with a 2-dimensional Gaussian function. The same in-plane indices of CFO and MSO were observed for samples less than 10 nm in thickness. Furthermore, lattice relaxation was observed in samples with thicknesses exceeding 10 nm.
\afterpage{\newpage}
\newpage
\clearpage
Figure \ref{fig:Mtplot} displays the CFO thickness $t$-dependence of the areal saturation magnetization $M_{\mathrm{S}}\cdot t$ for Samples A-1, A-2, and CFO/MgO. Additionally, linear fitting lines were shown in the Figure \ref{fig:Mtplot}. In case of CFO/MgO, the intercept of the fitting line is in the positive region of the horizontal axis, indicating the presence of a magnetic dead layer with a thickness of 3 $\pm$ 0.5 nm. The dead layer can be attributed to the introduction of antiphase boundaries caused by the difference in the crystal structures between the CFO and MgO structures\cite{VANDERHEIJDEN1996,BALL1996-1,Ball1996-2,Eerenstein2002,Moussy2004,Ramos2006}. In CFO/MSO, the intercept of the horizontal axis is -0.1 $\pm$ 0.4 nm and -0.8 $\pm$ 1.1 nm for Sample A-1 and A-2, respectively, indicating that the fitting line almost passes through the origin. Therefore, the dead layer thickness is negligible for CFO/MSO, implying that the formation of antiphase boundaries near the CFO interface is effectively suppressed by the use of a spinel MSO buffer. Figure \ref{fig:Ku_chi} shows the total distortion $\chi$-dependence of the induced $K_{\mathrm{u}}$ for Sample A-1 and Sample A-2. The max value of $K_{\mathbf{u}}$ was 6.1 MJ~m$^{-3}$.

\afterpage{\newpage}
\newpage
\clearpage

\section*{Torque analysis (45$^{\circ}$ method)}
\begin{figure}[h]
    \centering
    \includegraphics[width=\linewidth]{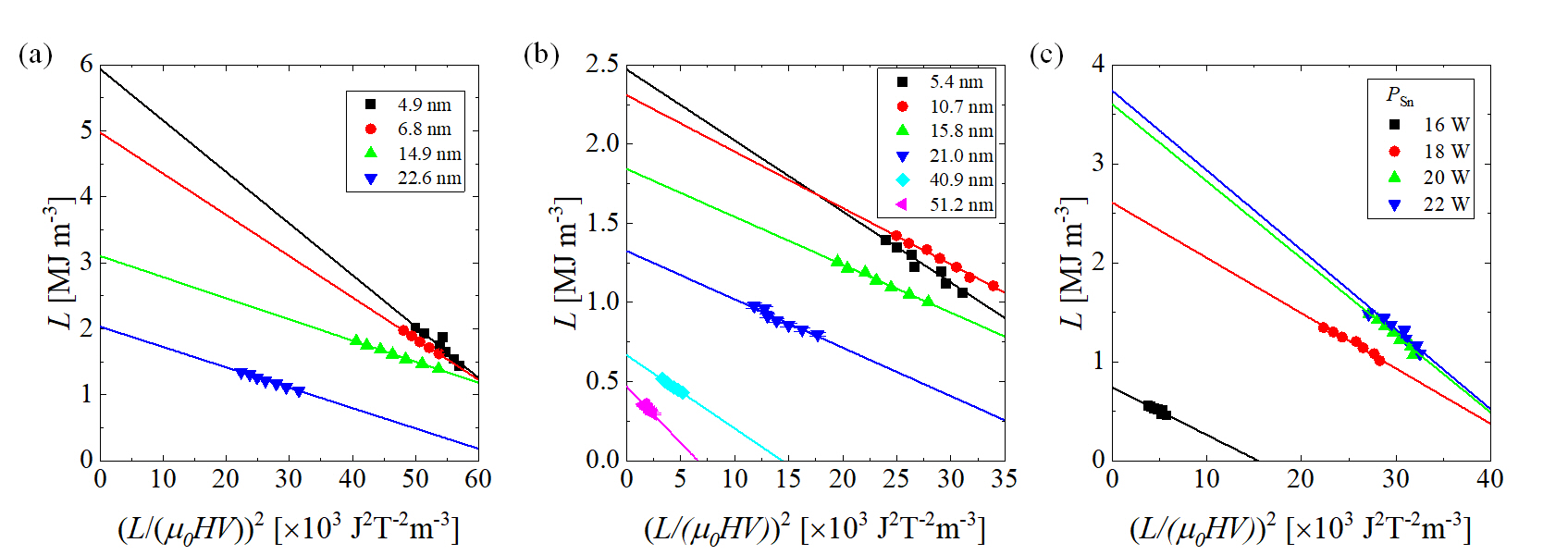}
    \caption{45$^{\circ}$ plot for (a) Sample A-1, (b) Sample A-2, and (c) the samples prepared in Method B.}
    \label{fig:Miyajima}
\end{figure}

When the anisotropy field of the sample is large, the 45$^{\circ}$ method proves effective, wherein the torque $L$ is plotted as a function of $(L/(\mu_0HV))^2$ at a fixed angle. When plotting the magnetic torque $ L $ that is measured under the applied field H in the direction of 45$^{\circ}$ to the film surface, the plot becomes linear. Several examples of 45$^{\circ}$ method are given in Figure \ref{fig:Miyajima}.  
\afterpage{\newpage}

\end{document}